# Spatial and Temporal Extrapolation of Disdrometer Size Distributions Based on a Lagrangian Trajectory Model of Falling Rain

John E. Lane[*]
ASRC Aerospace
Space Life Sciences Lab
Kennedy Space Center, Florida 32899, USA
john.e.lane@nasa.gov

Takis Kasparis
School of Electrical Engineering and Computer Science
University of Central Florida, Orlando, Florida 32816, USA
kasparis@ucf.edu

Philip T. Metzger
NASA/KSC Granular Mechanics and Regolith Operations Lab
Kennedy Space Center, Florida 32899, USA
philip.t.metzger@nasa.gov

W. Linwood Jones
School of Electrical Engineering and Computer Science
University of Central Florida, Orlando, Florida 32816, USA
ljones5@cfl.rr.com

**ABSTRACT:** Methodologies to improve disdrometer processing, loosely based on mathematical techniques common to the field of particle flow and fluid mechanics, are examined and tested. The inclusion of advection and vertical wind field estimates appears to produce significantly improved results in a Lagrangian hydrometeor trajectory model, in spite of very strict assumptions of noninteracting hydrometeors, constant vertical air velocity, and time independent advection during a radar scan time interval. Wind field data can be extracted from each radar elevation scan by plotting and analyzing reflectivity contours over the disdrometer site and by collecting the radar radial velocity data to obtain estimates of advection. Specific regions of disdrometer spectra (drop size versus time) often exhibit strong gravitational sorting signatures, from which estimates of vertical velocity can be extracted. These independent wind field estimates can be used as initial conditions to the Lagrangian trajectory simulation of falling hydrometeors.

[*] Corresponding Author

# INTRODUCTION

Weather radar measures the backscatter in terms of reflectivity $Z$, of an ensemble of instantaneously suspended raindrops in a volume defined by a microwave beam range increment and beam width solid angle. Mechanical rain gauges sample a related ensemble of drops at the surface and at a later time due to advection and hydrometeor terminal velocities. Rainfall rate $R$ is measured at the ground by the volume of rainwater accumulating in a collector per unit time. The most direct method of obtaining a relationship between these two types of measurements is to compare rainfall at the gauge to the collocated radar reflectivity. However, point rainfall measurements are seldom well correlated to the corresponding volume radar reflectivity measurements. This may be due to the large discrepancies in the sampling volumes, hydrometeor evaporation, gravitational sorting of drops, advection, updraft/downdraft velocity, uncertainties in drop terminal velocity, and contrasting spatial and temporal sampling resolutions.

The physical connection between rainfall rate $R$ measured by rain gauges and $R(Z)$ estimated by weather radar is the drop size distribution (DSD), and both can be described by integrals involving the DSD. The quality and characteristics of comparisons between $Z$ and ground-based measurements of $R$ are strongly influenced by the existing type and configuration of gauge instrumentation. Practical considerations have led the National Weather Service to use a single Z-R relation based on the power law, $Z=AR^b$, with constant $\{A,b\}$ parameters regardless of possible spatial or temporal variations. These same issues pertain to the measurement of hail but with an additional complication. Rain is commonly found without hail, but hail is seldom seen without rain. Therefore, when the goal is to make accurate measurements on hail size distributions, rain tends to corrupt those measurements.

Many observable properties of an ensemble of rain and hail can be explained by a simplified model of hydrometeor dynamics. Instrumentation that measures these properties involves either measurements of some volume attribute of the ensemble or characteristics of the flux at a surface. Rain gauges measure the flux accumulation or rate of accumulations at the ground. Disdrometers record the number and size of individual particles. Weather radar measures the microwave backscatter of an ensemble of hydrometeors as characterized by a drop size distribution. Hydrometeor size distributions are a function of space and time, where the size, state, and shape determine the still air terminal velocity. No single instrument can measure all properties of a hydrometeor ensemble – each instrument measures some aspect of the total set of properties. Merging data from multiple instruments, such as rain gauges, disdrometers, and radar, supplemented by a physical model of the size distribution and dynamics, provides the best possible view of the time varying, spatially dependent ensemble of hydrometeors that we call precipitation.

# HYDROMETEOR MEASUREMENT AND ANALYSIS

Most precipitation reaching the ground in the tropics does so in the form of raindrops and occasionally hail. From a very simplistic point of view, these hydrometeors can be viewed as noninteracting and falling at a constant terminal velocity. Even though this assumption is not generally accurate, because of numerous physical processes which govern drop dynamics, such as breakup, evaporation, and coalescence, it is nevertheless a useful assumption for the purpose of analyzing many important properties of rainfall. A one-way coupled two-phase system approach for computing the dynamics of particles in a fluid system is standard practice when two-way coupling is not practical. For example, the interaction of aerosol droplets in a compressed gas stream may be, to first order, predicted by modeling the trajectories of individual noninteracting droplets propelled by the gas that is unaffected by the presence of the droplets. A similar methodology can be used to approach analysis of hydrometeor trajectories, which may be highly influenced by the fluid motion of the ambient air.

The convention adopted in this work to describe the ambient air motion is to imply a cylindrical coordinate system so that wind vectors at every point in space are composed of horizontal and vertical components. Keeping with the meteorological convention, the horizontal wind component is referred to as the *advection* velocity. The vertical component is then referred to as the *updraft* velocity for a positive vertical movement, or *downdraft* velocity for a negative vertical motion.

**Disdrometer Gravitation Sorting Signature**

The primary motivation for this work at the start was the observation and subsequent attempt to model gravitational sorting of raindrops in disdrometer spectra. Gravitational sorting observations in disdrometer data have been reported in the literature on numerous occasions, but it is not a common observation simply because of the way in which disdrometer data is typically stored and plotted. The *gravitational sorting* signature is best observed when every drop impact measured by the disdrometer is time tagged and then displayed as a scatter plot of drop diameter *D* vs. time *t*. The resulting *D-t* plots often show marked diagonal features, where these gravitational sorting signatures are characterized by a negative slope when plotted as *D* vs. *t*. This can be explained by a simplistic model of noninteracting drops traveling at terminal velocity $v_D$ as a function of drop diameter *D*, along with an ambient vertical wind motion *w*. At a time $t = t_0$, a pulse of rain characterized by a typical DSD, such as the Marshall Palmer [1] or gamma distribution [2], starts at a height *h* above the ground site where a disdrometer is acquiring raindrop spectra (i.e., counting and measuring drop sizes). The height *h* is an idealized point of rainfall generation where a rainfall DSD, such as the MP DSD at $z = h$, completely determines the size distributions of falling drops. This point should be correlated to a level within a cloud where rainfall begins its unimpeded trek to the ground, such as the 0° C isotherm.

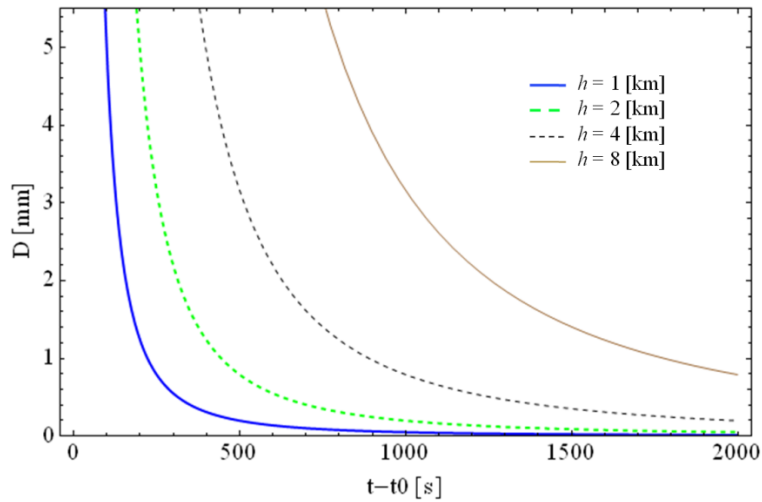

**Fig. (1).** Plots of *D(t)* due to idealized pulse rain characterized by a delta function, using Eq. (2) for various values of *h* with $w = 0$.

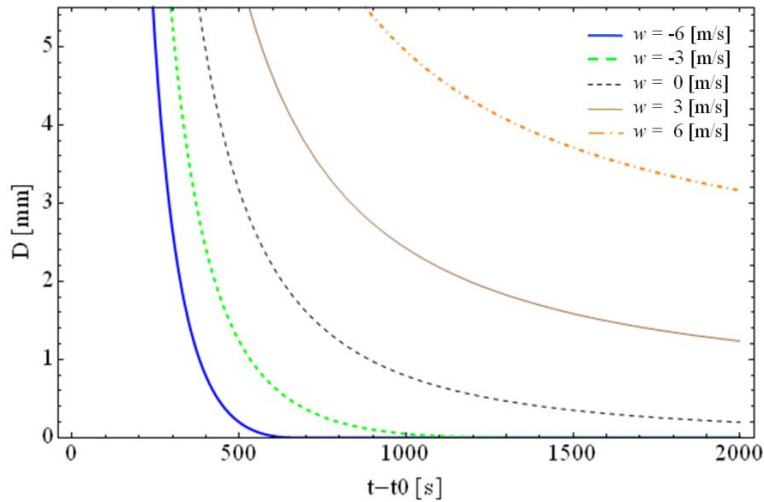

**Fig. (2).** Plots of *D(t)* due to idealized pulse rain characterized by a delta function, using Eq. (2) for various values of *w* with $h = 2000$ m.

Other hydrometeor phase types can be treated similarly, but some difficulties with mixed phases (rain, hail, etc.) may be encountered because of factors such as overlap in size distributions (small hail may be smaller than large rain drops); differences in terminal velocities (round hail versus flattened rain drops); and differences in radar reflectivity (ice has a somewhat higher radar reflectivity than rain for equivalent scattering cross sections).

The gravitational sorting disdrometer signature can be modeled based on the simplifying assumption of noninteracting drops by first expressing the hydrometeor fall time as:

$$\tau_D = h / (v_D - w) \tag{1}$$

where $\tau_D = t_D - t_0$ is the fall time of a hydrometeor of diameter $D$; $t_D$ is the clock time of the disdrometer recorded hit; $v_D$ is the terminal velocity; and $w$ is the vertical air motion. An estimate of terminal velocity for drop sizes associated with normal rain, is $v_D \approx aD^b$, where $a \approx 4.5$ m s$^{-1}$ mm$^{-b}$, and $b \approx 1/2$ [3,4]. Substituting this expression into Eq. (1) and solving for the drop diameter $D$, results in:

$$D = \begin{cases} \left(\dfrac{h/\tau_D + w}{a}\right)^{1/b} & h/\tau_D + w > 0 \\ undefined & h/\tau_D + w \leq 0 \end{cases} \tag{2}$$

Figs. (**1 – 2**) show Eq. (2) plotted for various values of $h$ and $w$. Eq. (2) is an idealized description of gravitational sorting raindrops due to a zero width rain pulse. Fig. (**3**) is a Monte Carlo simulation of raindrops generated by 60 s pulse width MP DSD. The simulated disdrometer data of Fig. (**3**) differs from the real disdrometer data of Fig. (**4**) for several reasons, reasons that will become more obvious in subsequent sections below.

The vertical feature shown in the $h = 0$ case on the left side of Fig. (**3**), can also occur for $w \to -\infty$, where the raindrops generated at $h$ and $t = t_0$ are swept downward by an infinite velocity downdraft to appear immediately at the ground to be recorded by the disdrometer. Neither the $h = 0$ or $w \to -\infty$ cases are physical cases, but serve as a convenient means to test and exercise the simulations.

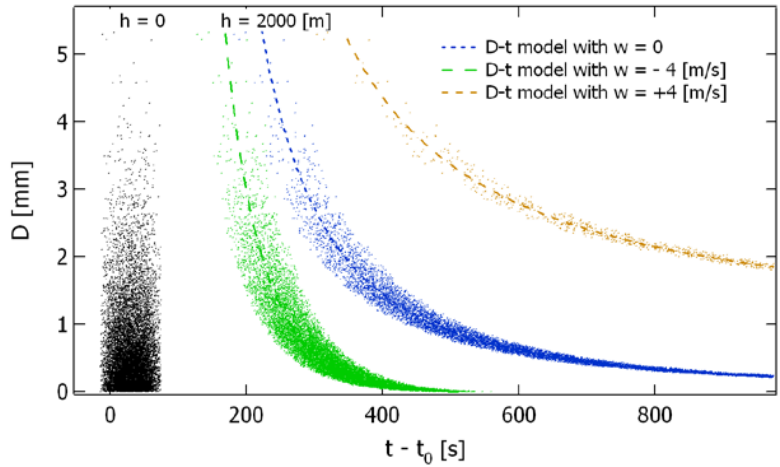

**Fig. (3).** Simulated disdrometer data for a $\tau = 60$ s pulse of $R = 100$ mm h$^{-1}$ rainfall using $10^6$ Monte Carlo drops generated with an MP drop size distribution.

Pulsed rainfall is commonly found in convective cells and thunderstorms where large variations of the time dependence of rainfall rate are common, where the most extreme case is a square pulse in both time and space. The gravitational sorting signature is less common in the case of stratiform rainfall because the time derivative of rainfall rate is small, so that no pulse-like conditions are observed. The $D$-$t$ character shown on the left side of Fig. (**3**) (the $h = 0$ case), occurs commonly in real disdrometer data when the rainfall rate is constant. It is also somewhat common to see the sharp discontinuities of *D-t* as shown in this simulation. This feature can occur during convective rainfall or stratiform rain due to advection. Advection causes the rainfall source, which is of finite extent, such as a convective cell, to pass over the disdrometer. In that case, the sharpness of the start and stop of the *D-t* scatter plot is due to the sharpness of the spatial extent of the rainfall source.

Fig. (**4**) shows disdrometer data from the University of Central Florida (UCF) Joss disdrometer site in Orlando, Florida (lat: 28.6016, lon: -81.1986), corresponding to September 30, 2008, from 19:00 – 21:00 universal time coordinated (UTC) time. Even though the individual drops impacts are not time tagged in the Joss data, the stored histogram size interval and time interval are sufficiently small to generate a *D-t* scatter plot. The count in the histogram drop size-time interval is plotted as a *density plot* in Fig. (**4**) such that a higher count in the histogram bin appears as a darker spot on the plot. Near the end of the rainfall event, gravitational sorting features due to pulsed rain become clearly visible. Eq. (2) is manually fitted to several of these features and overlaid on the plot.

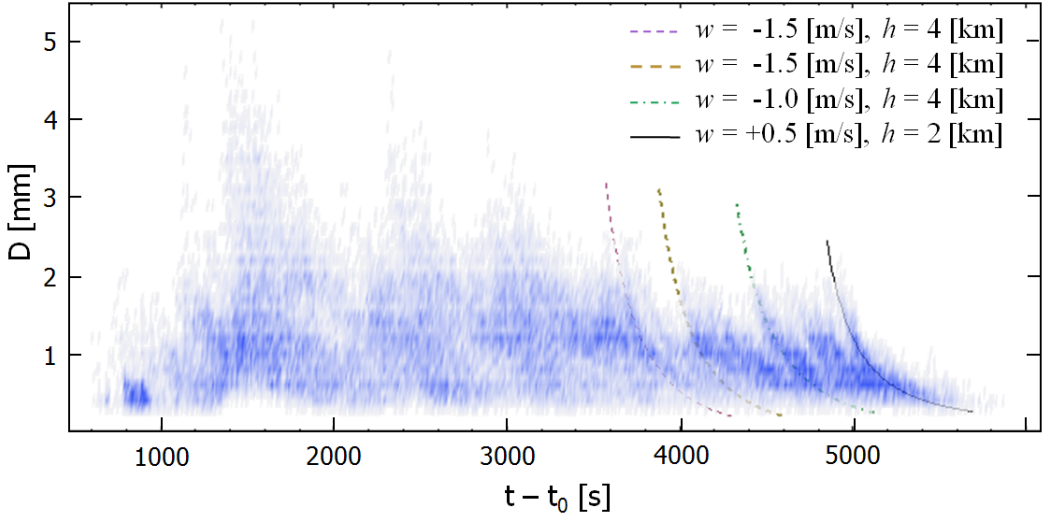

**Fig. (4).** Joss disdrometer data from the University of Central Florida site, September 30, 2008, with $t_0$ = 19:00 UTC. *D(t)* plotted is from Eq. (2).

**Disdrometer Derived Rainfall Products**

From the disdrometer histogram, an equivalent drop size distribution, rainfall rate, and disdrometer computed radar reflectivity can be calculated [5]:

$$N_{jk} = \frac{H_{jk}}{v_{D_j} A_S \Delta t \Delta D} \qquad \textit{Drop Size Distribution} \qquad (3)$$

$$R_k = \frac{\pi}{6 A_S \Delta t} \sum_{j=1}^{M} D_j^3 H_{jk} \qquad \textit{Rainfall Rate} \qquad (4)$$

$$Z_k = \frac{1}{A_S \Delta t} \sum_{j=1}^{M} \frac{D_j^6}{v_{D_j}} H_{jk} \qquad \textit{Radar Reflectivity} \qquad (5)$$

where $H_{jk}$ is the Joss disdrometer generated histogram corresponding to elapsed time $k\Delta t$ and drop size $D_j$; $A_S$ is the Joss sensor area equal to 50 cm$^2$; and $v_{D_j}$ is the terminal velocity corresponding to drop diameter size $D_j$. In Fig. (**4**) $\Delta t = 10$ s and the histogram has been resampled to a uniform diameter size so that $\Delta D = 0.1$ mm and $D_j = j\Delta D$. In Fig. (**5**), $\Delta t = 30$ s and $D_j$ is based on the Joss table of $M = 127$ nonuniformly spaced diameter sizes.

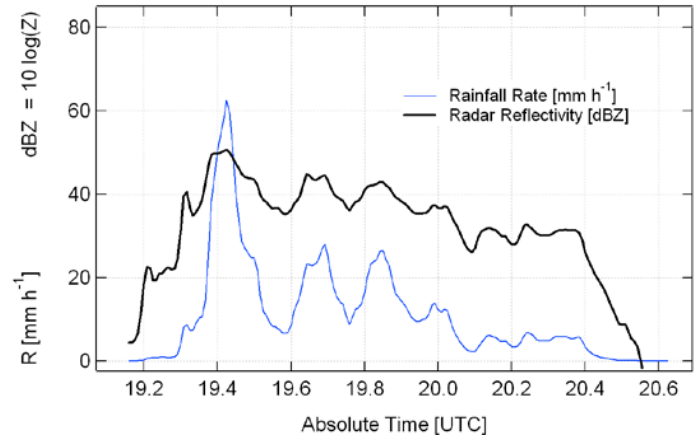

**Fig. 5.** Rainfall products derived from September 30, 2008, Joss disdrometer at the University of Central Florida: thin line is rainfall rate computed by Eq. (4); thick line is equivalent radar reflectivity computed using Eq. (5).

**Radar Data**

The Melbourne National Weather Service (NWS) radar is located approximately 55 km to the southeast of the University of Central Florida Joss site. For purposes of this study, radar data in only the lowest four scan elevations will be processed, and only in a 2×2 km horizontal extent centered over the Joss site. This region is later expanded to 8×8 km in order to extract the advection velocity. For the majority of this analysis, the radar reflectivity data is plotted in a pseudo 3D format, using *3DRadPlot* [6]. This display format was adopted and utilized in the study of hail events surrounding the Shuttle Launch Pad 39 structures at the Kennedy Space Center [7]. The same format is useful in analyzing the UCF Joss disdrometer data.

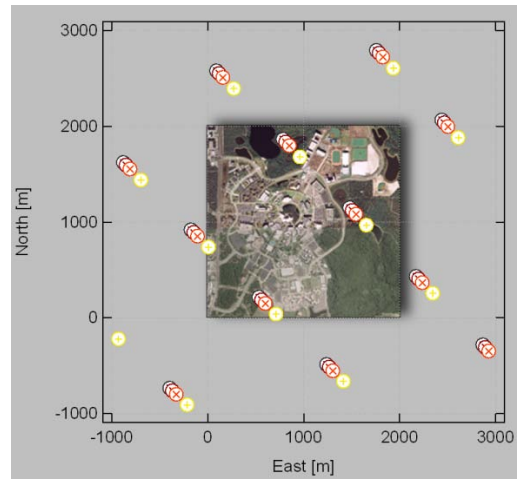

**Fig. (6).** UCF disdrometer site (Joss disdrometer in located in center). Circles represent locations of the Melbourne NWS radar bins for the lowest four elevations (09-30-08): black circles, $z = 1.00$ km, $t = 19:24:06$ UTC; brown circles, $z = 2.40$ km, $t = 19:24:41$ UTC; red circles, $z = 3.50$ km, $t = 19:25:23$ UTC; and yellow circles, $z = 4.90$ km, $t = 19:25:45$ UTC.

The basis of *3DRadPlot* display processing is built on a simplified and customized version of Shepard's interpolation formula [8]. For any point in space, the interpolated reflectivity $Z(\mathbf{r})$ at $\mathbf{r} = \{x, y, z\}$ is due to the weighted average of all reflectivity data points $Z_i$ at the center of all radar bin locations $\mathbf{r}_i$ in the local region:

$$Z(\mathbf{r}) = \frac{\sum_{i=1}^{N} Z_i |\mathbf{r} - \mathbf{r}_i|^{-p}}{|\mathbf{r} - \mathbf{r}_i|^{-p}} \tag{6}$$

where $N$ is the number of reflectivity data points used in the interpolation, for example, $N = 48$, the number of circles in Fig. (**6**), the number of radar bins in the region defined by 4×4 km × 4 scan elevations. The exponent parameter $p$ controls the rate of transition between actual data points and the interpolated values. The larger this value, the sharper the transition between actual values defined by $\mathbf{r}_i$. In this work, a value of $p = 8$ was used throughout. Figs. (**7-8**) are examples of *3DRadPlot* output plots of reflectivity for 19:23 UTC and 19:38 UTC respectively. This can be compared to the reflectivity derived from the Joss as shown in Fig. (**5**).

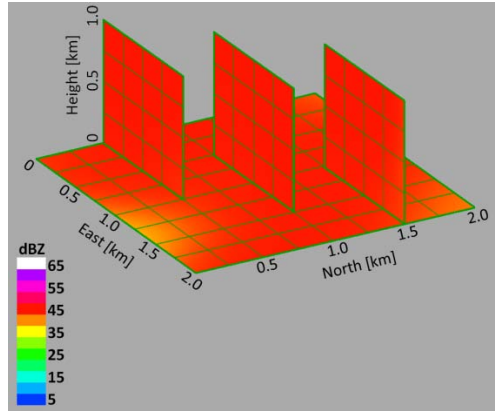

**Fig. (7).** *3DRadPlot* of Melbourne radar reflectivity over UCF Joss site, 09-30-08, 19:23 UTC.

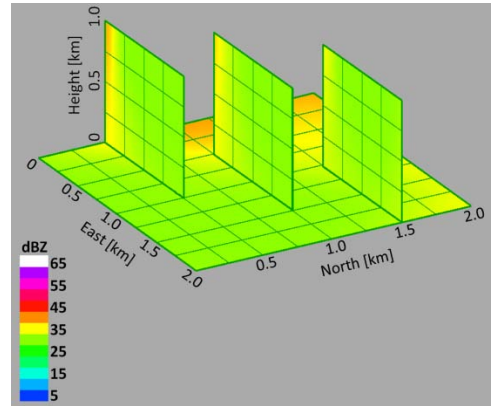

**Fig. (8).** *3DRadPlot* of Melbourne radar reflectivity over UCF Joss site, 09-30-08, 19:38 UTC.

**Joss Derived Products**

Fig. (**9**) is a Joss histogram corresponding to Fig. (**4**), summed over all $k$ time intervals. This in itself is not a useful product, but it is easy to generate and may serve as a convenient method of categorization of rainfall events. The formal drop size distribution specified by Eq. (3) is a much more useful quantity in comparing rainfall events,

especially since it is normalized by time. Fig. (**10**) is the Joss derived drop size distribution using Eq. (3), for the entire storm event of Fig. (**4**), with $\Delta t = 6000$ s. The MP DSD equivalent corresponds to an average rainfall rate of 10 mm h$^{-1}$. Since the length of the storm event may be ambiguous, it is preferable to define shorter time intervals and then generate multiple DSDs as a function of time. Another common method is to group the DSDs by rainfall rate as opposed to grouping by time.

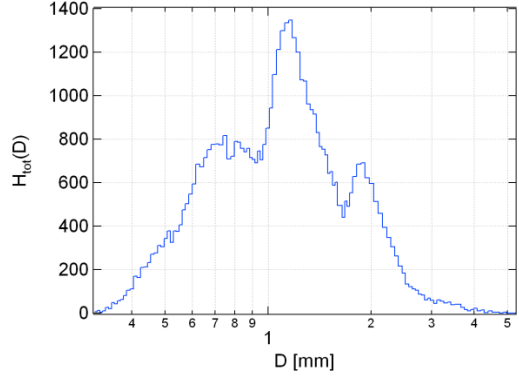

**Fig. (9).** Joss histogram corresponding to Fig. (**4**), summed over all $k$ time intervals.

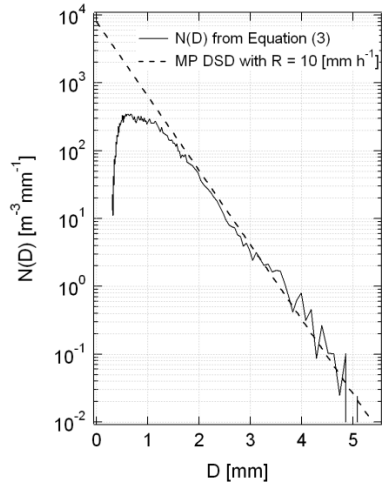

**Fig. (10).** Joss derived drop size distribution using Eq. (3), for entire storm event of Fig. (**4**), with $\Delta t = 6000$ s. The MP DSD equivalent corresponds to an average rainfall rate of 10 mm h$^{-1}$.

Fig. (**11**) is a comparison of the Joss derived rainfall rate from Eq. (4) $\Delta t = 30$ s (thick line) and the rainfall rate from three collocated rain gauges (thin lines). The Joss and three rain gauges are clustered together within a 5 m diameter area on the roof of the UCF Engineering I Building in Orlando, Florida. The total storm accumulated gauge rainfall average is 19.39 mm, as compared to the 16.26 mm derived rainfall rate. The cause of this 16% difference in rainfall accumulation between the Joss and rain gauges is unknown. Suspected causes may be interference from wind on the roof of the building under conditions of high rainfall rate or a disdrometer that is in need of calibration. This discrepancy is larger than it should be, but it is not the topic of this work and will not interfere significantly with the results presented in this paper. In future work we would hope to have a better initial correlation between disdrometer derived rainfall rate and collocated rain gauge data.

Fig. (**12**) is close-up of disdrometer derived reflectivity for the lowest four elevation scans over the UCF site, also referring to Fig. (**6**), with time marks for 19:23 UTC and 19:38 UTC scans. The average disdrometer $Z$ for 19:23 UTC is 50.1 dBZ and 43.9 dBZ for 19:38 UTC when averaging over the two sets of four elevation scan time marks. The radar reflectivity of Fig. (**7**) for 19:23 UTC is approximately 45 dBZ, about 5 dBZ lower than the disdrometer derived value. The radar reflectivity shown in Fig. (**8**) for 19:38 UTC is approximately 30 dBZ or about 14 dBZ lower than the disdrometer derived value. Proposing and testing explanations for these kind of discrepancies are the major goals of this work. The remainder of the material presented in this paper addresses this issue. The solution strategy takes into account the gravitational sorting observations in the disdrometer data. It also should be noted at this time that final tactics for improving the disdrometer derived radar reflectivity will involve strategies that will ultimately compute the disdrometer extrapolated 4D-DSD (a function of *x*, *y*, *z*, *t*, and *D*). Then, all disdrometer products can be computed directly from the 4D-DSD, including the equivalent radar reflectivity based on the sixth moment of the 4D-DSD.

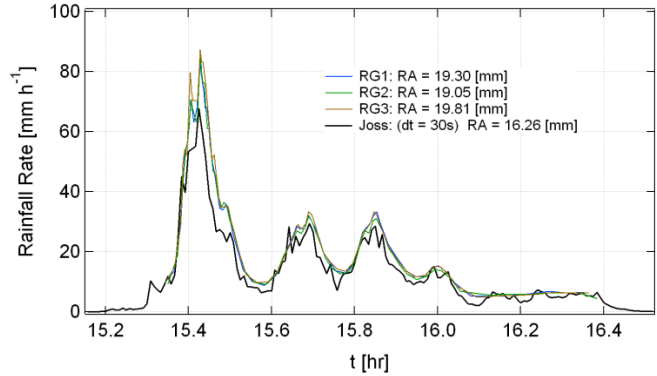

**Fig. (11).** Thick line is rainfall rate derived from the Joss data using Eq. (4) with $\Delta t = 30$ s. The three thin lines are rainfall rate measured by three collocated rain gauges. The total storm accumulated gauge rainfall average is 19.39 mm, as compared to the 16.26 mm for the Joss.

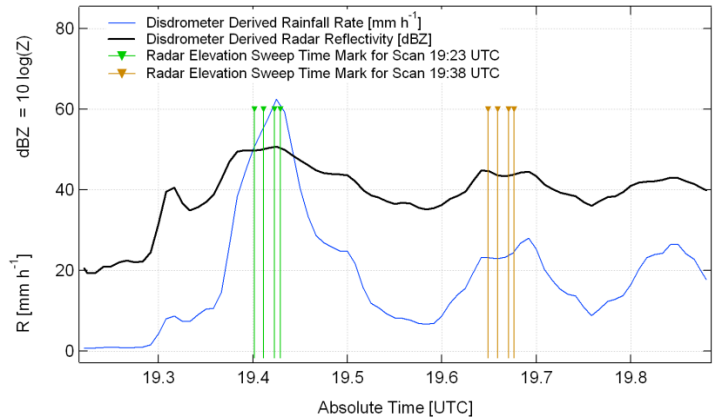

**Fig. (12).** Close-up of disdrometer derived reflectivity corresponding to the lowest four elevation scans over the UCF site, referring also to Fig. (**6**), for 19:23 UTC and 19:38 UTC scans with time marks. The average *Z* for 19:23 UTC is 50.1 dBZ and 43.9 dBZ for 19:38 UTC. This should be compared to the reflectivity values shown by the NEXRAD data of Figs. (**7-8**).

**Incorporation of Drop Fall Time**

Fig. (**13**) depict three methodologies for computing disdrometer derived reflectivity. The first method in Fig. (**13a**) is simply the direct application of Eq. (4), as was demonstrated in the previous section for the 19:23 UTC radar scan, in which case a 5 dBZ discrepancy was observed between the disdrometer and radar reflectivities. Fig. (**13b**) illustrates an improvement to this simple case in which a time delay is applied to compensate for the average fall

time from the height of the center of the radar bin based on an average drop terminal velocity. This intermediate case is presented without an example because our real interest is the third case of Fig. (**13c**). This is the *gravitational sorting time delay* case where the disdrometer time delay is based on gravitational sorting from Eq. (1) and is dependent on the terminal velocity $v_D$ of each drop size as well as the vertical air motion $w$. The examples to be examined will now be based on comparing the simple case of Fig. (**13a**) and two cases from Fig. (**13c**), one with $w = 0$ and one with $w$ chosen to enforce a good comparison (in a least squares sense) between the disdrometer $Z$ and the radar $Z$. In order to incorporate Eq. (1), the computation of disdrometer derived reflectivity, from Eq. (5), must be modified by letting $k'\Delta t = k\Delta t + Int[\tau_D]$, resulting in:

$$Z_k = \frac{1}{A_S \Delta t} \sum_{\substack{j=1 \\ v_{D_j} > w}}^{M} \frac{D_j^6}{v_{D_j}} H_{jk'} \tag{7a}$$

$$k' = k + \begin{cases} Int\left[\dfrac{z}{(v_{D_j} - w)\Delta t}\right] & v_{D_j} > w \\ undefined & v_{D_j} \leq w \end{cases} \tag{7b}$$

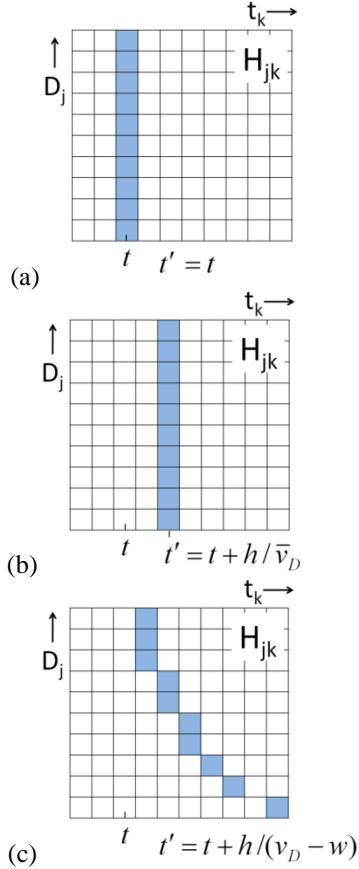

(a)

(b)    $t' = t + h/\bar{v}_D$

(c)    $t' = t + h/(v_D - w)$

**Fig. (13).** Comparison strategies for radar reflectivity and disdrometer derived reflectivity: (**a**) *simple* case where the time $t$ of the radar scan is matched to the corresponding disdrometer data at $t = t'$; (**b**) the *simple time delay* case where $Z(t)$ is matched to the disdrometer data delayed by the average drop terminal velocity $\bar{v}_D$ falling from a height $z = h$, (**c**) *gravitational sorting time delay* case where the time delay is dependent on the terminal velocity $v_D$ of each drop size as well as vertical air velocity $w$.

The presence of $v_D$ in Eq. (7b) implies that the index $k'$ in Eq. (7a) is a function of index $j$, which depends on the details of the terminal velocity approximation used. Note the vertical air velocity is not used in Eq. (7a) since it is assumed that the vertical air motion is zero near the ground and up to the height that is required to reach terminal velocity. For the largest drop sizes, it is assumed that vertical air motion is zero within a vertical region extending to 10 to 20 m above the disdrometer. Otherwise, Eq. (7a) would need to be modified to account for the effect of vertical air movement on drop velocities at the disdrometer.

The summation must be constrained on several fronts. To examine this in more detail, we will need to pick an explicit form of $v_D$. Using the drop terminal velocity from a previous section, $v_D \approx aD^b$, and replacing $D$ with the discrete value $D_j = j\Delta D$, results in:

$$k' = k + \begin{cases} Int\left[\dfrac{z}{(a\Delta D^b j^b - w)\Delta t}\right] & j > j_0 \\ undefined & j \leq j_0 \end{cases} \qquad (8)$$

Eq. (7a) may now be expressed in terms of the drop size index $j$ as:

$$Z_k = \frac{1}{A_s a \Delta t} \sum_{j=j_0}^{M} \frac{(j\Delta D)^6}{(j\Delta D)^b} H_{jk'}$$
$$= \frac{\Delta D^{6-b}}{A_s a \Delta t} \sum_{j=j_0}^{M} j^{6-b} H_{jk'} \qquad (9a)$$

where,

$$j_0 = 1 + \begin{cases} Int\left[\dfrac{w^{\frac{1}{b}}}{a^{\frac{1}{b}}\Delta D}\right] & w > 0 \\ 0 & w \leq 0 \end{cases} \qquad (9b)$$

and since the sum in Eq. (9a) is truncated for $j \leq j_0$, the time delayed index $k'$ can be written as:

$$k' = k + Int\left[\frac{z}{(a\Delta D^b j^b - w)\Delta t}\right] \qquad (9c)$$

Eqs. (9) define an algorithm to compute a disdrometer derived reflectivity that accounts for drop fall time due to terminal velocity as well as vertical air motion, corresponding to the case depicted in Fig. (**13c**). Eqs. (9) can be forced to reduce to the simple case of Fig. (**13a**) and Eq. (5), by letting $w \to -\infty$. The result is that $w \to -\infty$ and $j_0 \to 1$, so that Eq. (9a) reduces to Eq. (5). Physically, this would represent a hypothetical infinite downdraft which transports the drops to the ground at the disdrometer in zero time, so that the radar reflectivity then corresponds to the disdrometer derived reflectivity. This absurd hypothetical case highlights the equally absurd basis of applying the simple case of Fig. (**13a**) and Eq. (5) for comparing disdrometer reflectivity to radar reflectivity, a common practice in radar meteorology work.

An additional limit that needs to be noted is due to the finite length of $H_{jk}$ in the $k$ direction (discrete time axis). $k'$ from Eq. (9c) can exceed this limit under several conditions. When this occurs, the summation in Eq. (9c) must simply be truncated or $H_{jk}$ set to zero for those values of $k'$.

Table 1 compares the UCF Joss disdrometer derived reflectivity corresponding to consecutive radar scans from 19:04 – 19:38 UTC, on September 30, 2008. The *3DRadPlot* format is utilized for easy comparison of the volume scan reflectivity in a 2×2×1 km box centered over the Joss disdrometer. Note that because of the geometry, only the lowest elevation scan will have a significant impact on the displayed reflectivity. Even so, all points in the lowest four scans are used from Fig. (**6**) in the Shepard interpolation. Recall that the disdrometer data is resampled to a constant drop size bin width, with $\Delta D = 0.1$ mm, and the original sample time, $\Delta t = 10$ s are used in Eqs. (9).

**Table 1. Application of Eqs. (9) to compute disdrometer derived $Z$ with comparison to Melbourne equivalent radar $Z$ on a scan by scan basis, using UCF Joss data of September 30, 2008.**

| Disdrometer derived reflectivity $Z$ using the simple case of Fig (**13a**) and Eq. (5). | Disdrometer $Z$ using method of Figure 13c and Eqs (9) with $w = 0$. | Disdrometer $Z$ using method of Fig. (**13c**) and Eqs. (9) with manual optimization of $w$. | Melbourne NEXRAD reflectivity plotted in a 2×2×1 km volume centered over Joss site. |
|---|---|---|---|
| 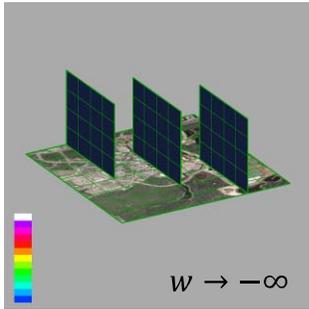 $w \rightarrow -\infty$ | 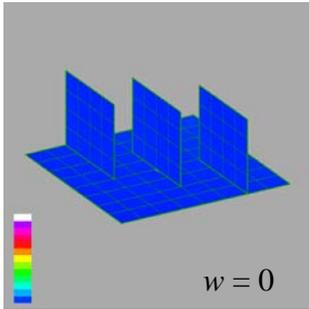 $w = 0$ | 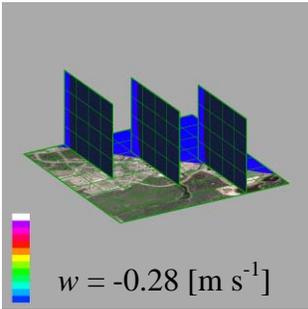 $w = -0.28$ [m s$^{-1}$] | 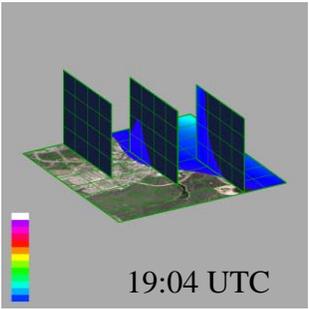 19:04 UTC |
| 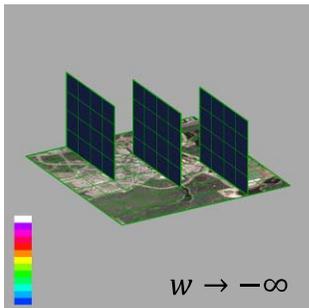 $w \rightarrow -\infty$ | 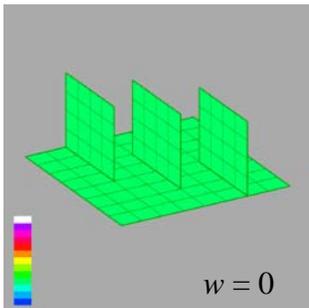 $w = 0$ | 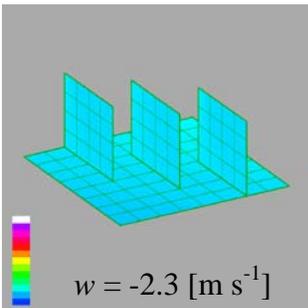 $w = -2.3$ [m s$^{-1}$] | 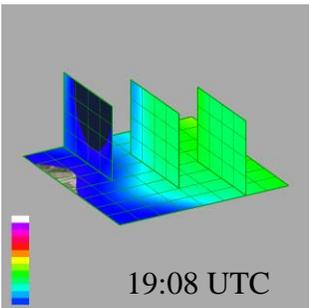 19:08 UTC |
| 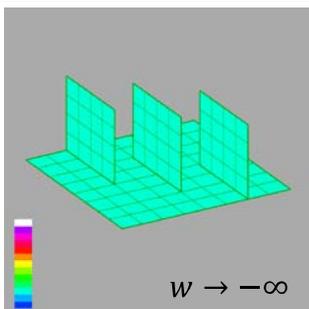 $w \rightarrow -\infty$ | 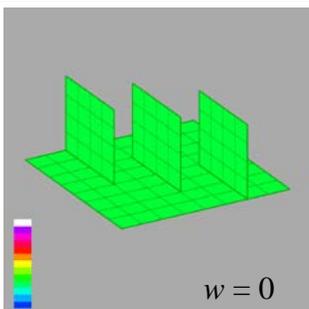 $w = 0$ | 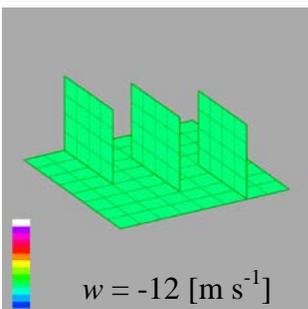 $w = -12$ [m s$^{-1}$] | 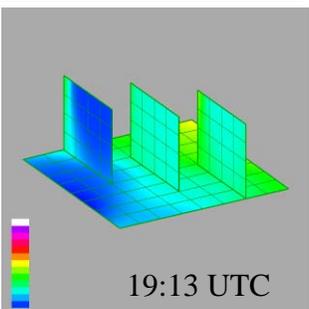 19:13 UTC |

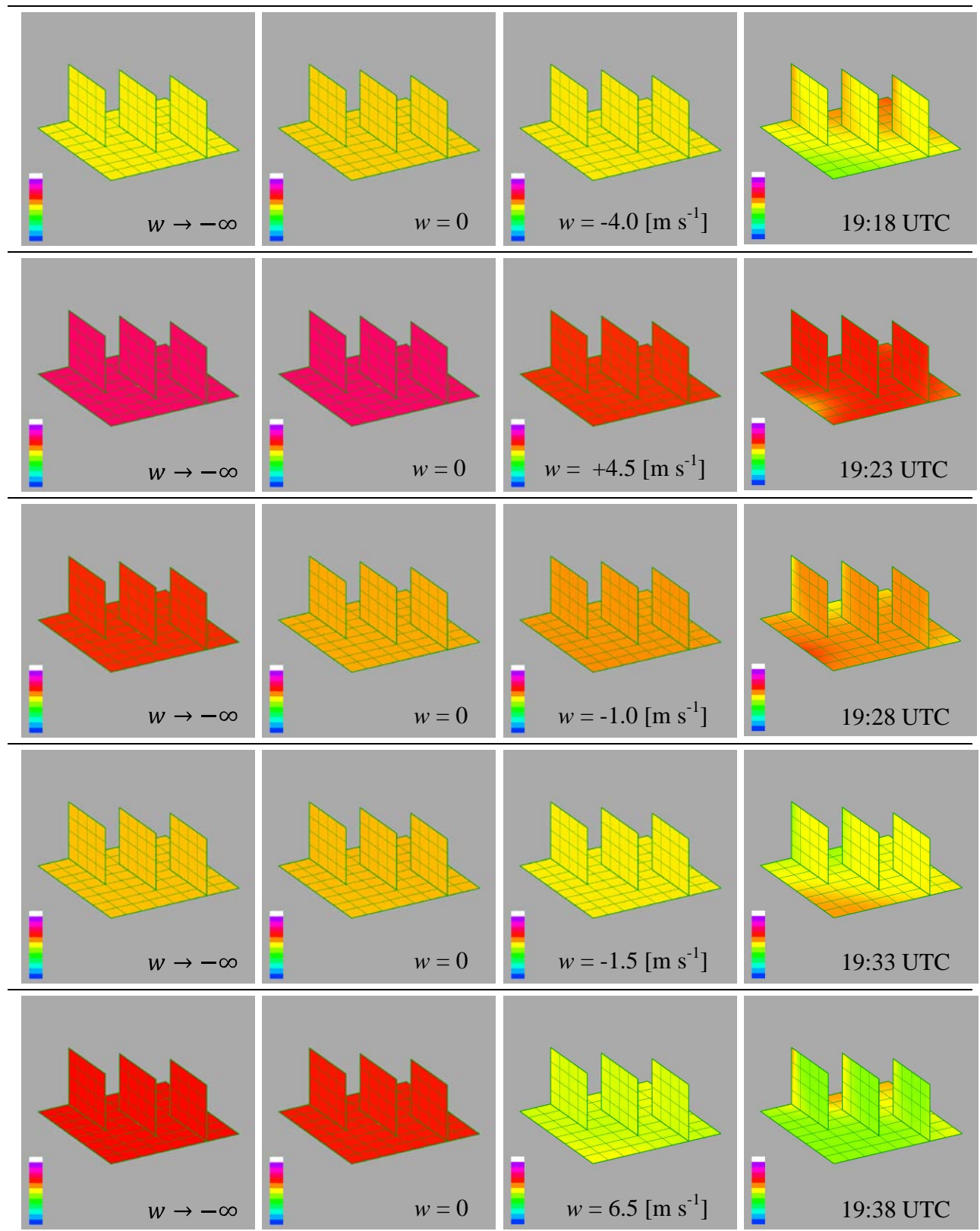

In order to obtain a more meaningful comparison with radar, disdrometer Z is computed via Shepard's interpolation at all of the radar bin points shown in Fig. (**6**), with the actual height values $z$ and time of scan defined by index $k$. Note that the exact positions of those points are contained in the Level III product data file and change position from volume scan to volume scan. The first column is the disdrometer Z using the simple case of Fig. (**13a**) and Eq. (5); however, it is calculated by the numerically approximation of setting $w = -999$ in Eq. (9). The second column is computed similarly as the first column but with $w = 0$. The third column is computed by choosing a $w$ that gives the best (very subjective) overall match to the radar data of the last column. In general, there is an improvement from left to right. In some cases, for example, the 19:38 UTC scan, the first two columns are drastically off from the radar data, but an empirical guess of $w = 6.5$ m s$^{-1}$ results in a good comparison.

Consideration of horizontal advection as well as drop vertical fall time provides additional improvement in the calculation of disdrometer reflectivity, as will be shown in the following section.

**Incorporation of Cloud Advection Effects**

In this approach, radar reflectivity and radial velocity data are used in addition to disdrometer spectra, where the end goal is no longer just an improved comparison between radar Z and disdrometer Z. The goal is to arrive at a good prediction of the DSD as a function of all three spatial coordinates, as well as time. Once that is accomplished, the resulting 4D-DSD can then be used to calculate rainfall products, analogous to Eqs. (3-5) for the one-dimensional case. Nevertheless, it is still convenient and useful to compare the final disdrometer derived reflectivity from the sixth moment of the 4D-DSD to the corresponding radar reflectivity.

Even though there are typically four dozen radar bin locations available at each complete scan in the 2×2 km area surrounding the UCF disdrometer site as shown in Fig. 6, far fewer than that number contribute significantly to the estimate of the 4D-DSD, primarily those in the lowest elevation scan. Based on Fig. (**6**), there are a half dozen especially influential points. Therefore, the number of free and arbitrary model fitting parameters should not be allowed to exceed the number of radar reflectivity points in order to properly satisfy least squares fitting requirements. However, this is more of an intelligent guideline than a strict rule since the approach taken in this work is not to implement a stringent least squares comparison of radar and disdrometer reflectivities.

Table 2 shows the Melbourne weather radar radial velocity over the disdrometer site for several scans of the example rainfall event for the first three elevations. This data was acquired using the NOAA Weather and Climate Toolkit, version 2.2, from the National Climatic Data Center, a free software download [9]. Note that both *classified* (velocity quantized to multiple of 5 kts) and *unclassified* data are shown in Table 2. The important characteristic to note in this data is the sometimes strong vertical gradient of the horizontal winds. This feature is most notable at the beginning of the storm where many atmospheric processes may interact, such as gust fronts and sea breeze collisions, as well as strong updrafts and downdrafts. On the backside of the storm (last two rows of Table 2), the vertical gradient is small or nonexistent.

An empirical advection model that is both relevant to the 4D-DSD volume and consistent with the observations of vertical advection gradients in the NEXRAD radial velocity, was used in this work:

$$\mathbf{u}(z) = \left(\gamma + (1-\gamma)\tanh\frac{z-z_0}{L_z}\right)\mathbf{u}_0 \tag{10a}$$

with,

$$\mathbf{u}_0 = u_0 \begin{pmatrix} \cos\psi \\ \sin\psi \end{pmatrix} \tag{10b}$$

The model described by Eqs. (10) exhibits two asymptotic values: $\mathbf{u}(z) \to (2\gamma-1)\mathbf{u}_0$ for $(z-z_0)/L_z \ll 0$ and $\mathbf{u}(z) \to \mathbf{u}_0$ for $(z-z_0)/L_z \gg 0$. The center of the transition at $z = z_0$ is $\mathbf{u}(z) = \gamma\mathbf{u}_0$ and the rate of the transition between asymptotic values is controlled by the parameter $L_z$. Note that $z_0$ is simply a fitting parameter and has no significant physical meaning. The 4D-DSD model is implemented in a Cartesian coordinate system with positive *x* pointing east, positive *y* pointing north, and positive *z* pointing up. Therefore, advection angle $\psi$ adheres to the polar angle convention of the Cartesian system so that $\psi = 0$ simulates an advection travelling from west to east; $\psi = 90°$ simulates an advection travelling from south to north, and so on.

**Table 2. Level II Melbourne NEXRAD radial velocity data over the UCF Joss disdrometer site, September 30, 2008.**

| *t* UTC | Elev deg | *z* m | Classified $V_r$ m/s | Unclassified $V_r$ m/s |
|---|---|---|---|---|
| 19:04 | 0.47 | 959  | -       | -2.52056 |
|       | 1.44 | 2246 | -       | -1.49176 |
|       | 2.39 | 3511 | -10.288 | -5.50408 |
| 19:08 | 0.47 | 960  | -5.144  | -4.47528 |
|       | 1.44 | 2246 | 5.144   | 1.49176  |
|       | 2.39 | 3513 | -5.144  | -1.49176 |
| 19:18 | 0.47 | 960  | -10.288 | -6.01848 |
|       | 1.44 | 2247 | 5.144   | 0.97736  |
|       | 2.39 | 3512 | 9.7736  | 2.52056  |
| 19:23 | 0.47 | 962  | -10.288 | -8.02464 |
|       | 1.44 | 2245 | -5.144  | -2.52056 |
|       | 2.39 | 3515 | 5.144   | 0.5144   |
| 19:28 | 0.47 | 963  | -10.288 | -6.99584 |
|       | 1.44 | 2246 | 5.144   | 1.49176  |
|       | 2.39 | 3508 | 5.144   | 0.97736  |
| 20:22 | 0.47 | 963  | -5.144  | -4.47528 |
|       | 1.43 | 2243 | -5.144  | -4.98968 |
|       | 2.39 | 3513 | -5.144  | -3.49792 |
| 20:27 | 0.47 | 959  | -5.144  | -4.01232 |
|       | 1.43 | 2242 | -5.144  | -4.47528 |
|       | 2.39 | 3515 | -5.144  | -4.01232 |

A simulated radial velocity can be calculated from Eq. (10a) by performing the vector dot product with the radar direction vector:

$$U_r(z) = \begin{pmatrix} \cos\theta\cos\varphi \\ \cos\theta\sin\varphi \\ \sin\theta \end{pmatrix} \cdot \begin{pmatrix} \mathbf{u}(z) \\ w \end{pmatrix} \tag{11a}$$

$$\theta = \theta_0 + \frac{s}{\frac{4}{3}R_E} \tag{11b}$$

where $\phi$ is the direction from the radar to the disdrometer site (in the Cartesian coordinate system of the 4D-DSD model); $\theta$ is the local radar scan elevation angle above the disdrometer; $\theta_0$ is the radar scan elevation angle relative to the radar site; *s* is the distance along the earth's surface from radar site to disdrometer site; and $R_E$ is the standard average earth radius, equal to 6371 km. The last term in Eq. (11b) takes into account the curvature of the earth, while the factor of 4/3 accounts for radar refraction to first order. Eq. (11b) assumes that the radar and disdrometer sites are both at points on a perfect spherical earth of radius $4R_E/3$.

Since under most conditions, the vertical component of $U_r(z)$ is very small compared to the horizontal component, setting $\theta = 0$ simplifies the simulated radial velocity. Combining Eqs. (10) and (11) with $\theta = 0$ yields:

$$U_r(z) \approx u_r(z) = u_0 \left[ \gamma + (1-\gamma)\tanh\frac{z-z_0}{L_z} \right] \cos(\varphi - \psi) \tag{12}$$

In order to compare the model radial velocities from Eq. (12) to the NEXRAD radial velocities $V_r$, such as those in Table 2, a vertical integration needs to be performed over the radar beam height at each elevation angle:

$$\begin{aligned}\bar{u}_r(m) &= \frac{1}{z_{m2}-z_{m1}}\int_{z_{m1}}^{z_{m2}} u_r(z) \\ &= u_0 \left[ \gamma + (1-\gamma)\frac{L_z}{z_{m2}-z_{m1}}\log\left(\cosh\frac{z_0-z_{m2}}{L_z}\cdot\operatorname{sech}\frac{z_0-z_{m1}}{L_z}\right)\right]\cos(\varphi-\psi)\end{aligned} \tag{13}$$

where $z_{m1}$ and $z_{m2}$ are the bottom and top edges of the beam at the *m*th radar beam elevation angle.

An average advection velocity $\bar{\mathbf{u}}(m)$ for the *m*th radar beam elevation angle can be estimated by plotting equal dBZ contours between consecutive scans, such that the contour lines bracket the disdrometer site. It is necessary to use the appropriate contour, corresponding to the dBZ value passing over the ground site at each scan time. For example, Fig. (14) shows the NEXRAD 5 dBZ reflectivity contour for the $m = 1$ scan (lowest elevation scan) centered over the disdrometer site for the 19:04 and 19:08 UTC scans. The procedure depicted in Fig. (14) can be repeated for all elevation angles of interest and for all time scans of interest.

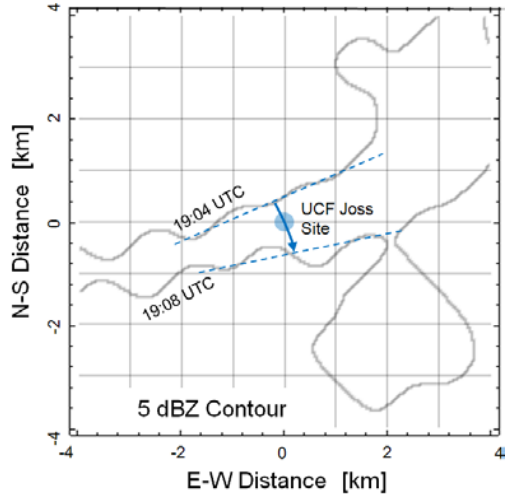

**Fig. (14).** Lowest elevation angle plot of the Melbourne NEXRAD 5 dBZ reflectivity contour over the UCF disdrometer site. Based on the translation of the equal contour line, the corresponding advection velocity over the site, using the notation of Eq. (13), is approximately $\bar{u}_c(1) \approx 3.3\,\text{m s}^{-1}$, with $\psi = -68°$.

Once the Z-contour velocities $\bar{\mathbf{u}}_c(m)$ have been estimated from radar reflectivity, for at least $m = 1$, self consistency of the radar data can be checked by comparing the radial components of $\bar{\mathbf{u}}_c(m)$ with the Doppler radial velocities, such as those shown in Table 2. The radial component of $\bar{\mathbf{u}}_c(m)$ is found by taking the dot product with the direction vector from the radar, similar to Eqs. (11). This comparison provides some confidence of the integrity of the radar derived advection velocity. In some cases, the Doppler radial velocity may be unreliable because of spatial and temporal fluctuations. The Z-contour derived advection may be unreliable in other cases, such as during collisions of the sea breeze and frontal boundary movement, in which case advection may be undefined. For the purpose of disdrometer spatial and temporal extrapolation of drop size distributions, $\bar{\mathbf{u}}_c(m)$ is be a better measure of relevant advection. For example, during times that a convective cell is stationary, $\bar{\mathbf{u}}_c(m)$ is a better indicator that advection is zero or undefined, than the associated Doppler velocities which may be nonzero. And most importantly, the Doppler velocity provides only the radial component of advection.

The goal of the discussion of the last several paragraphs is to find a method to choose the best parameters for the advection model, Eqs. (10), based on empirical matching to the Doppler and reflectivity derived horizontal air motion at the first two or three elevation angles. For example, the advection model corresponding to the 19:04 UTC radar scan is shown in Fig. (**15**). It should be noted that the advection model of Eq. (10a) only has the ability to change the velocity magnitude, not the direction. It can, however, reverse the sign. Increasing the capabilities is of the advection model is straight-forward and a primary candidate for future work.

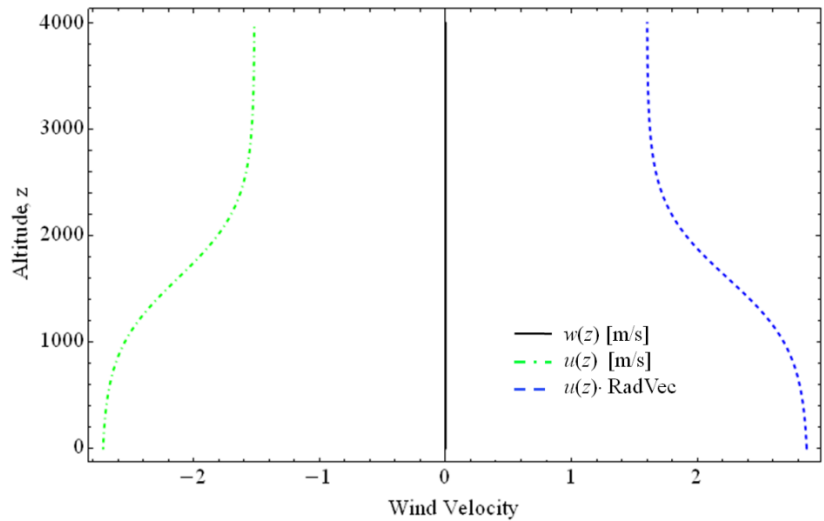

**Fig. (15).** Advection model of Eqs. (10) for scan 19:04 UTC, with: $\gamma = 1.4$; $L_z = 700$ m; $z_0 = 1600$ m; $\psi = -64°$, $u_0 = 1.6$ m s$^{-1}$; and $\phi = 134.5°$.

**GENERALIZED LAGRANGIAN RAINFALL MODEL**

The previous sections described independent horizontal and vertical trajectory models of falling hydrometeors due to gravity and the effects of ambient air motion. The primary effect of atmospheric air movement is due to drag forces that either move the particle along with the ambient air and/or limit the drop fall velocity, i.e. terminal velocity. Other effects include particle lift due to the rotation of the particle or rotation of the air around the particle. Lift forces, which may be positive or negative, are always in the direction of the gravity vector and are proportional to the difference in horizontal components of particle motion and air velocity. Lift is usually much smaller in magnitude than drag forces, except in the case of very fast rotations. Mass loss or gain is another effect that controls the detailed motion of hydrometeors. Evaporation, collisions, spontaneous breakup, and coalescence are mechanisms that further complicate drop dynamics.

A traditional approach to simulating trajectories of single particles in a gas flow is to employ recursive integration [10], such as the fourth order Runge-Kutta algorithm (RK4) [11] to compute the particle position at each time step. Note that there is a typo in Eq. (25.5.20) of ref. [11] for $k4$: the term $hk_3/2$ should be $hk_2/2$. The inputs to the integration algorithm include local gas velocity vector, density, viscosity, and temperature at each of the particle's time-stepped positions. For some applications such as in aeronautics, these gas properties might be the output from a *computation fluid dynamics* (CFD) or *direct simulation Monte Carlo* (DSMC) model of the gas flow.

Advanced fluid mechanics models utilize a two-way coupled system where the gas flow is affected by the particle flow, and the particle movement is in turn affected by the gas flow. Less computationally intensive software systems implement a one-way coupled model where the gas flow affects the particle flow, but the particle flow has no affect on the gas flow.

The horizontal advection and vertical wind components of falling hydrometeors can be combined in a one-way coupled algorithm, using physical property formulas of the surrounding air as input to the Lagrangian trajectory integration. For a hydrometeor of diameter $D$ and mass $m$, the trajectory is due to the external forces: gravity, vertical updrafts/downdrafts, and advection. The sum of external forces on a hydrometeor is equal to its acceleration, which can be estimated by a second order Taylor series (TS2) expansion about time point $n$, resulting in a set of difference equations for position and velocity [10]:

$$\mathbf{v}_n = \mathbf{v}_{n-1} + \mathbf{a}_{n-1} \Delta t \tag{14a}$$

$$\mathbf{r}_n = \mathbf{r}_{n-1} + \mathbf{v}_{n-1} \Delta t + \tfrac{1}{2} \mathbf{a}_{n-1} \Delta t^2 \tag{14b}$$

$$\mathbf{U}_n = \mathbf{U}(\mathbf{r}_n) = \begin{pmatrix} \mathbf{u}(\mathbf{r}_n) \\ w(\mathbf{r}_n) \end{pmatrix} \tag{14c}$$

$$\begin{aligned}
\mathbf{a}_n &= \frac{\pi C_D D^2}{8m} \left| \mathbf{U}_n - \mathbf{v}_n \right| \cdot \left( \mathbf{U}_n - \mathbf{v}_n \right) \rho_n + g_E \hat{\mathbf{e}}_E \\
&= \frac{\pi C_D D^2}{8} \left( \frac{\pi D^3 \rho_H}{6} \right)^{-1} \left| \mathbf{U}_n - \mathbf{v}_n \right| \cdot \left( \mathbf{U}_n - \mathbf{v}_n \right) \rho_n + g_E \hat{\mathbf{e}}_E \\
&= \frac{3 C_D \rho_n}{4 D \rho_H} \left| \mathbf{U}_n - \mathbf{v}_n \right| \cdot \left( \mathbf{U}_n - \mathbf{v}_n \right) + g_E \hat{\mathbf{e}}_E
\end{aligned} \tag{14d}$$

where $g_E$ is earth gravity (9.80665 m s$^{-2}$); $\rho_H$ is hydrometeor density (997.0479 kg m$^{-3}$ at 25 C); $\rho_n$ is the air density; $\mathbf{U}(\mathbf{r}_n)$ is the total air velocity at the $n$th time step particle position $\mathbf{r}_n$, and the acceleration $a_n$ is due to particle drag [12] and gravity. The direction of the gravity unit vector $\hat{\mathbf{e}}_E$ is given by:

$$\begin{aligned}
\hat{\mathbf{e}}_E &\equiv \frac{-1}{\sqrt{x^2 + y^2 + (R_E + x)^2}} \begin{pmatrix} x \\ y \\ R_E + z \end{pmatrix} \\
&\approx \begin{pmatrix} 0 \\ 0 \\ -1 \end{pmatrix} \qquad \text{for } x, y, z \ll R_E
\end{aligned} \tag{15}$$

Both the TS2 and RK4 methods can generate nearly identical results if the time step size is small enough. It is generally agreed that RK4 is far superior under conditions where the TS2 might fail. Both methods have been used in this work. Another note on notation convention: in this work, the convention used for recursive formulas is to show the present value being computed as indexed by $n$, which uses past values indexed by $n-k$, $k = 1, 2, \ldots$ The convention in [11], as with many other numerical methods, is to compute the $n+1$ value using $n-k$ values, with $k = 0, 1, \ldots$

The coefficient of drag, $C_D$ is a function of the Reynolds number, $Re$ [12]:

$$Re_n = \frac{D\rho_n |\mathbf{U}_n - \mathbf{v}_n|}{\mu_n} \tag{16}$$

where $\mu_n$ is the dynamic viscosity of air and is related to temperature using Sutherland's formula [13]:

$$\mu_n = \mu(\mathbf{r}_n) = \mu_0 \frac{T_0 + C}{T(\mathbf{r}_n) + C} \left( \frac{T(\mathbf{r}_n)}{T_0} \right)^{3/2} \tag{17}$$

where $\mu_0 \equiv 1.827 \times 10^{-5}$ Pa s, $T_0 \equiv 291.15$ K, and $C \equiv 120$ K. The air density $\rho_n$ in Eq. (14d) and (16) is related to air temperature and pressure using the well known *ideal gas law*:

$$\rho_n = \frac{P(\mathbf{r}_n) M_0}{R^* T(\mathbf{r}_n)} \tag{18a}$$

$$P(\mathbf{r}_n) = P_0 \left( 1 - \frac{\eta \, z_n}{T(0)} \right)^{\frac{g_E M_0}{R^* \eta}} \tag{18b}$$

$$T(\mathbf{r}_n) = T(0) - \eta \, z_n \tag{18c}$$

where $M_0$ is the molecular weight of dry air (0.0289644 kg mol$^{-1}$); $R^*$ is the gas constant (8.31432 J mol$^{-1}$ K$^{-1}$); $P_0$ is standard air pressure at sea level (101325 Pa); $\eta$ is the average rate of temperature change with altitude (0.0065 K m$^{-1}$); and $T(0)$ is the standard temperature at sea level (288.15 K). Eq. (18b) is a solution of the well known *equation of hydrostatic equilibrium*. Eq. (18c) is a solution to the *environmental lapse rate equation*, which is approximately linear in the troposphere [14].

The coefficient of drag can be computed from the following empirical formula [15]:

$$C_D = \begin{cases} 24.0 Re^{-1} & Re < 2 \\ 18.5 Re^{-0.6} & 2 \leq Re < 500 \\ 0.44 & Re \geq 500 \end{cases} \tag{19}$$

The initial conditions assume that the hydrometeor is at terminal fall velocity:

$$\mathbf{v}_0 = \begin{pmatrix} \mathbf{u}(\mathbf{r}_0) \\ w - v_D \end{pmatrix}, \quad \mathbf{r}_0 = \begin{pmatrix} x_0 \\ y_0 \\ z_0 \end{pmatrix}$$
(20)

where the drop terminal velocity $v_D$ is given by an approximation that accounts for altitude effects, such as Best [16].

Eqs. (14-20) define a complete Lagrangian trajectory model for hydrometeors, one that provides a method to integrate the equations of motion, following the hydrometeor path from an arbitrary starting point in space, $\mathbf{r}_0 = \{x_0, y_0, z_0\}$ to the ground. The time of travel $\tau = L \Delta t$ is found by noting the number of elapsed time steps $L$ when $\mathbf{r}_L \rightarrow \{x_L, y_L, 0\}$. The general strategy is to choose a set of starting points $\mathbf{r}_{0i}$, then run the trajectory Eqs.

(14–20) recursively until the hydrometeor encounters the ground, repeating this for all discrete values of drop diameter $D_j$ of the relevant drop size range. For every $D_j$ starting at $\mathbf{r}_{0i}$, the coordinates on the ground and the arrival time at the ground are logged. The times and locations of the hydrometeor ground strikes are then related to the disdrometer spectra by a time delay and spatial extrapolation from the disdrometer location to the coordinates of the hydrometeor strike. This procedure provides a method to generate a volume DSD in three dimensional space and in time. The Lagrangian approach to computing a 4D-DSD from the disdrometer data is a powerful method since it can easily accommodate complex spatial and temporal advection functions, as well as complex vertical wind profiles. In previous work, a 3D-DSD algorithm was implemented in software to process disdrometer data using a set of ballistic equations, which had been integrated analytically using simple functions of advection and vertical wind motion where the wind motion was assumed to be time independent over the period of drop trajectory calculations [17]. The Lagrangian trajectory method has no requirement of time independent wind motion. (Note that the difference in terminology adopted in this work is that previous work computed a 3D-DSD where time was involved only in the trajectory equations but was not used to compute changes in acceleration as a function of time).

**Disdrometer Derived Reflectivity**

A final problem to consider is that of spatial extrapolation from the disdrometer site to the point on the ground of the hydrometeor strike. An ad hoc approach is to assume that each $j$th-$k$th bin of the disdrometer histogram $H_{jk}$ travels from the disdrometer site to the hydrometeor impact point $\mathbf{p}_{ij} = \{x_{ij}, y_{ij}\}$ using the following transformation [17]:

$$t'_{ij} = t_i + \tau_{ij} - \frac{\mathbf{u}(0) \cdot \mathbf{p}_{ij}}{|\mathbf{u}(0)|^2} \tag{21}$$

where $\mathbf{u}(0)$ is the advection velocity near the ground at $z \to 0$; $\mathbf{p}_{ij}$ is the point of impact on the ground for the $j$th drop size falling from the $\mathbf{r}_{0i}$ position in space; $\tau_{ij}$ is the fall time of the $j$th drop size from the point $\mathbf{r}_{0i}$; and $t_i$ is the absolute time that drops begin to fall from the point $\mathbf{r}_{0i}$. The disdrometer data corresponding to $H_{jk'}$ is then used to estimate the extrapolated disdrometer spectra at the hydrometeor impact point $\mathbf{p}_{ij}$, where:

$$k' = Int\left[\frac{t'_{ij} - t_0}{\Delta t}\right] \tag{22}$$

and $t_0$ is the absolute start time of the disdrometer spectra acquisition that produces $H_{jk'}$. For the purpose of comparing disdrometer reflectivity to radar reflectivity, it is convenient to define the set of $\mathbf{r}_{0i}$ and $t_i$ corresponding to the set of radar bins over the disdrometer volume, as shown in Fig. (**6**).

In some cases, such as times when a sea breeze collides with a convective cell front, advection may not be well defined. In such cases where advection goes to zero or reverses direction, the last term on the right (the advection term) in Eq. (21) can be deleted, which essentially reduces to the case of Eq. (7b). In cases where advection is non-zero, but small or simply noisy, a linear combination of these two solutions can then be used to compute the drop size distribution:

$$N(D) = \beta\, N_1(D) + (1 - \beta)\, N_2(D) \tag{23}$$

where $\beta$ is an empirical parameter that mixes some percentage of the advective case $N_1(D)$ using the time delay defined by Eqs. (21-22) and the stationary nonadvective case, $N_2(D)$ from Eq. (7b).

**DISCUSSION AND SUMMARY**

Referring to the *3DRadPlot* graphics in Tables 1 and 3, the 19:04 and 19:08 scans show a distinct improvement in the disdrometer derived reflectivity when vertical fall time is incorporated into the processing algorithm (second and third columns compared to the first column). Calculations based on the Lagrangian-advection model, shown in the middle column of Table 3, also show good agreement with the radar in most cases. Scans 19:13, 19:18, and 19:23 all show good agreement in all columns with the corresponding radar reflectivity. One explanation for this behavior is that during the 19:13 – 19:23 period, rainfall is characterized by a very broad pulse, so that gravitational sorting is not a significant effect during that time interval (refer to the disdrometer spectra plot of Fig. (**4**) at 780 – 1380 s). In this case, the disdrometer reflectivity can more easily track the radar reflectivity.

Scans 19:28, 19:33, and 19:38 begin to show severe disagreement between disdrometer and radar. Only the processing methods shown in Table 3 display reasonable agreement with the radar reflectivity. An explanation for this case is that the rainfall is characterized by impulsive conditions (refer to Fig. (**4**) at 1680 – 2280 s). During this time, advection begins to degenerate due to interaction of the storm movement with surrounding opposing winds (this is somewhat apparent in the radar data). Another symptom of poorly defined advection is seen in Table 4 by examining the evolution of the mixing parameter $\beta$. Recall from Equation (23) that $\beta = 1$ is a purely advective solution, whereas $\beta = 0$ is a purely nonadvective solution defined by Eq. (7b). In Table 4, where all parameters are created by a manual process of finding a best fit to the corresponding radar plot, $\beta = 1$ during the nonimpulsive phase, then begins to decrease to smaller values. The only exception, during the 19:04 scan $\beta$ is less than 1, which may be explained by poor advection conditions at the earliest approach of the storm, when rainfall rate and reflectivity are very small. Another interesting feature in the parameter set of Table 4 is the evolution of the vertical velocity parameter *w*. Vertical velocity starts out at a small value or negative (downdraft), then begins to rise to larger and larger positive values (updraft), where the maximum coincides with the maximum radar reflectivity at 19:23.

Methods to improve disdrometer processing, loosely based on mathematical techniques common in the field of particle flow and fluid mechanics, have been explored. The inclusion of advection and vertical winds appears to produce significantly improved results, in spite of very strict assumption of noninteracting hydrometeors, constant vertical air velocity, and time independent advection during the scan time interval. Time dependent advection may be incorporated quite simply within the framework of the Lagrangian trajectory mechanics. Simulation improvements have been seen in modeling sprinkler systems by accounting for drop interactions [18]. Future work should focus on incorporating at the least, a simple model of drop interaction.

A most important activity of future work should be to exercise the model vigorously with a sufficient volume of data so that statistics of the model performance can be quantified. Along with this effort, all available independent wind field data should be rigorously collected and processed to be used as inputs into the Lagrangian model. Sources of this data may be derived directly from each radar elevation scan by plotting and analyzing reflectivity contours over the disdrometer site, such as that demonstrated in Fig. (**14**), and by collecting the radar radial velocity data. Strong gravitational sorting signatures in the disdrometer spectra are another potential source of vertical wind data. Other sources of data that would be very useful, if available, include colloacted anemometer or wind tower data. Rain gauge data from multiple gauges positioned within a kilometer or less of the disdrometer site would provide a valuable addition to the set of data required for 4D-DSD optimization and verification.

**Table 3.** Comparison of Lagrangian trajectory model disdrometer derived Z with Melbourne equivalent radar reflectivity on a scan by scan basis, using UCF Joss data of September 30, 2008. The first column is copied from the third column of Table 1.

| Disdrometer $Z$ using method of Fig. (**13c**) and Eqs. (9) with manual optimization of $w$. | Disdrometer $Z$ using Lagrangian trajectory model and Eqs. (21-23) with manual optimization of all parameters (see Table 4). | Melbourne NEXRAD reflectivity plotted in a 2×2×1 km volume centered over UCF Joss site. |
|---|---|---|
| 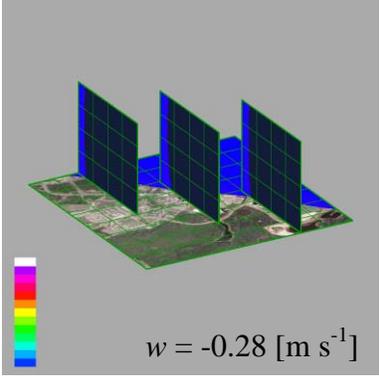 $w$ = -0.28 [m s$^{-1}$] | 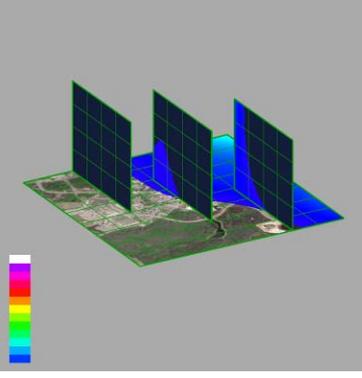 | 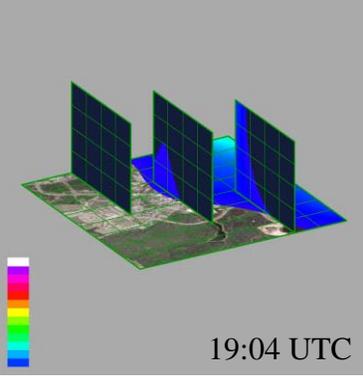 19:04 UTC |
| 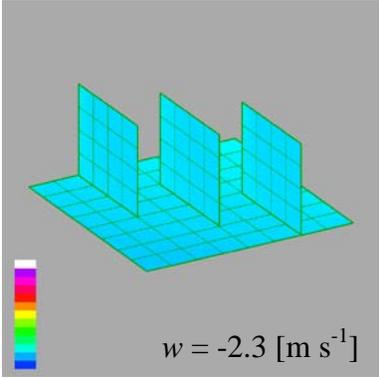 $w$ = -2.3 [m s$^{-1}$] | 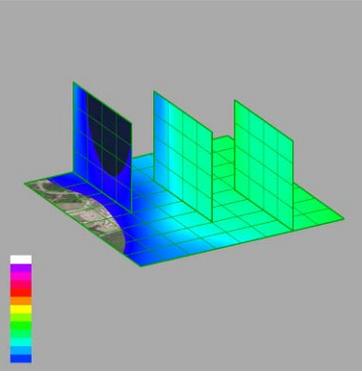 | 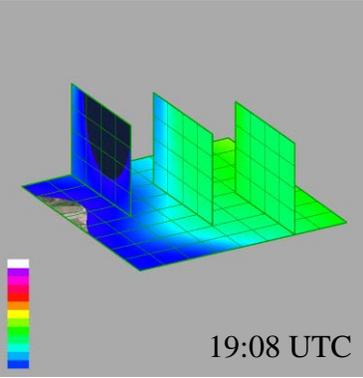 19:08 UTC |
| 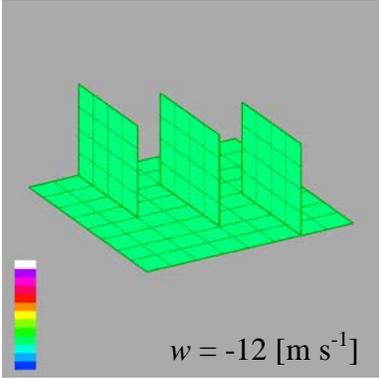 $w$ = -12 [m s$^{-1}$] | 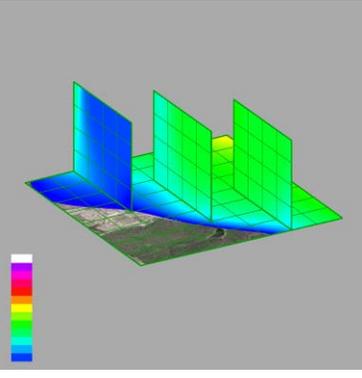 | 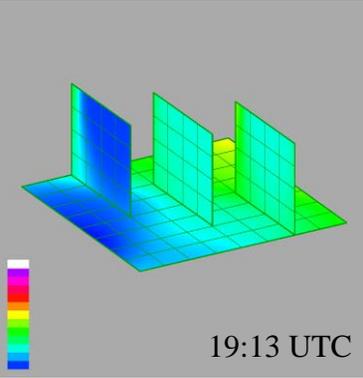 19:13 UTC |

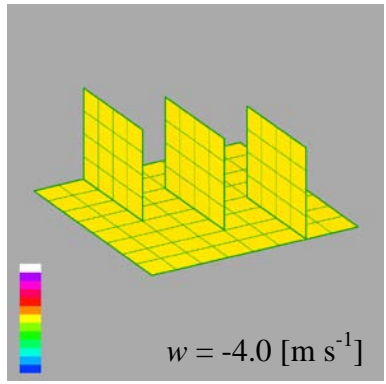 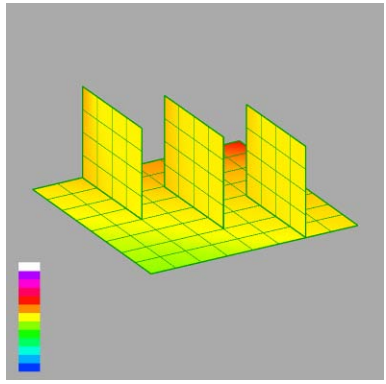 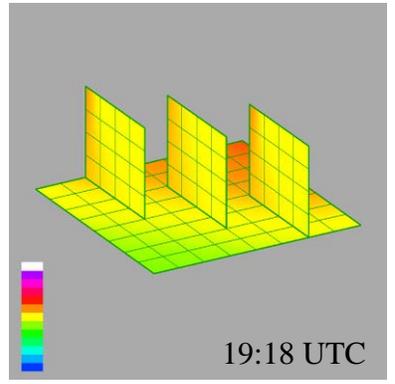

$w = -4.0$ [m s$^{-1}$]  19:18 UTC

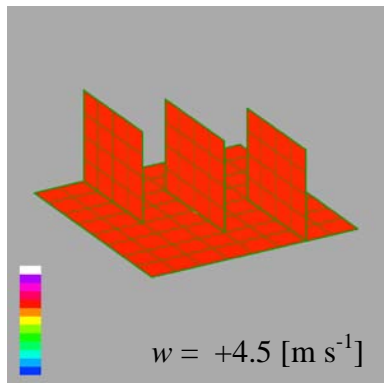 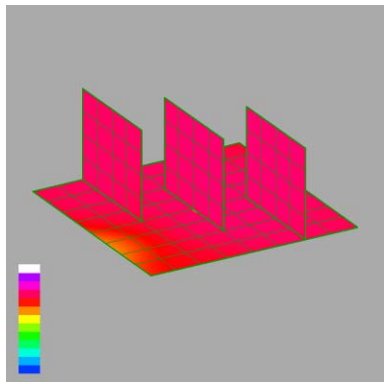 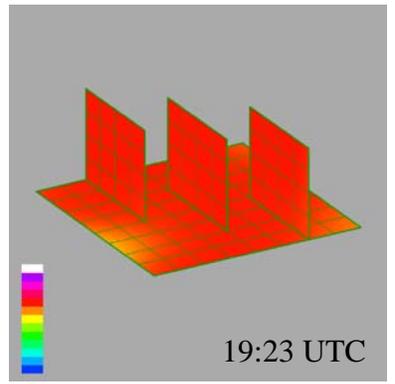

$w = +4.5$ [m s$^{-1}$]  19:23 UTC

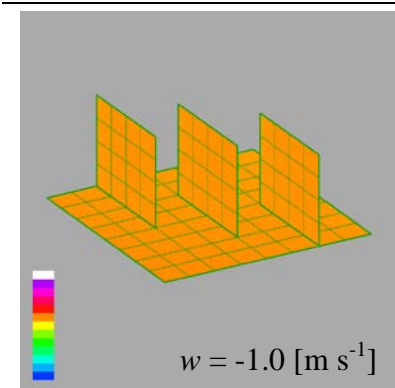 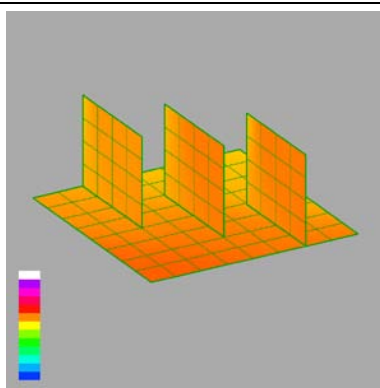 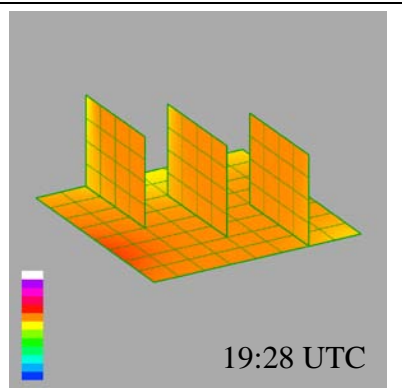

$w = -1.0$ [m s$^{-1}$]  19:28 UTC

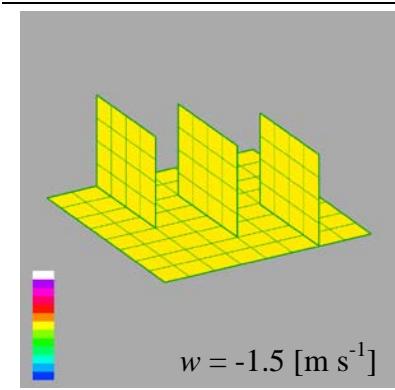 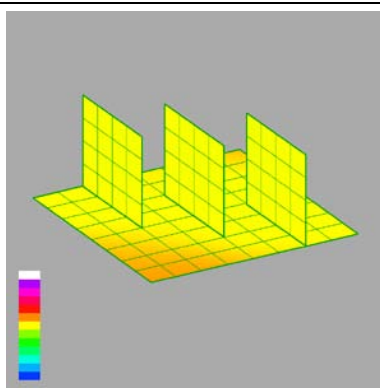 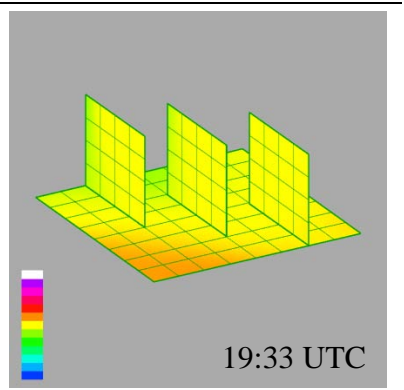

$w = -1.5$ [m s$^{-1}$]  19:33 UTC

**Table 4. Lagrangian model parameters used to produce second column in Table 3.**

| UTC Time | 19:04 | 19:08 | 19:13 | 19:18 | 19:23 | 19:28 | 19:33 |
|---|---|---|---|---|---|---|---|
| $u_0$ [m s$^{-1}$] | 1.6 | -1.3 | 6.2 | -1.8 | 1.35 | -1.6 | 5.5 |
| $\psi$ [deg] | -64 | -99 | -72 | -43 | -69 | -35 | -40 |
| $\beta$ | 0.61 | 1.0 | 1.0 | 1.0 | 1.0 | 0.55 | 0.29 |
| $\gamma$ | 1.4 | -4.0 | 0.52 | -7.7 | 4.2 | -7.0 | 1.0 |
| $L_z$ [m] | 700 | 920 | 350 | 800 | 850 | 600 | 500 |
| $z_0$ [m] | 1600 | 0 | 0 | 0 | 1500 | 600 | 0 |
| $w$ [m s$^{-1}$] | 0 | 0 | -2.5 | 1.0 | 4.5 | 4.0 | -5.7 |